\documentclass{aa}

\usepackage{graphicx}
\usepackage{txfonts}
\usepackage{natbib}
\bibpunct{(}{)}{;}{a}{}{,}

\begin{document}
  \title{Improved low-temperature rate constants for rotational
    excitation of CO by H${_2}$}
  
  \author{M. Wernli \inst{1} \and P. Valiron \inst{1} \and A. Faure
    \inst{1} \and L. Wiesenfeld \inst{1} \and P. Jankowski \inst{2}
    \and K. Szalewicz \inst{3}}
  
  \institute{Laboratoire d'Astrophysique de Grenoble, UMR 5571
    CNRS-UJF, OSUG, Universit\'e Joseph Fourier, BP 53, F-38041
    Grenoble Cedex 9, France \and Department of Quantum Chemistry,
    Institute of Chemistry, Nicholaus Copernicus University, Gagarina
    7, PL-87-100 Toru{\'n}, Poland \and Department of Physics and
    Astronomy, University of Delaware, Newark, DE 19716, USA }
  
  \date{Received 26 July 2005 / Accepted 12 September 2005}

  \offprints{M. Wernli \email{michael.wernli@obs.ujf-grenoble.fr}}

  \abstract{Cross sections for the rotational (de)excitation of CO by
  ground state para- and ortho-H$_2$ are obtained using quantum
  scattering calculations for collision energies between 1 and
  520~cm$^{-1}$. A new CO$-$H$_2$ potential energy surface is employed
  and its quality is assessed by comparison with explicitly correlated
  CCSD(T)-R12 calculations.  Rate constants for rotational levels of
  CO up to 5 and temperatures in the range 5$-$70~K are deduced. The
  new potential is found to have a strong influence on the resonance
  structure of the cross sections at very low collision energies. As a
  result, the present rates at 10~K differ by up to 50$\%$ with those
  obtained by \citet{flower01} on a previous, less accurate, potential
  energy surface.

  \keywords{molecular data $-$ molecular processes $-$ ISM: molecules}
  }
  
  \titlerunning{improved low-temperature CO$-$H$_2$ rotational rate
  constants} \maketitle

%

\section{Introduction}

Since its discovery in interstellar space more than 30 years ago
\citep{wilson70}, carbon monoxide (CO) has been extensively observed
both in our own and in other galaxies. CO is indeed the second most
abundant molecule in the Universe after H$_2$ and, thanks to its
dipole moment, it is the principal molecule used to map the molecular
gas in galactic and extragalactic sources. CO is also currently the
only molecule to have been observed at redshifts as high as $z=6.42$
\citep{walter03}. Collisional excitation of rotational levels of CO
occurs in a great variety of physical conditions and emission from
levels with rotational quantum numbers $J$ up to 39 have been
identified in circumstellar environments \citep{pepe96}.
 
Hydrogen molecules are generally the most abundant colliding partners
in the interstellar medium, although collisions with H, He and
electrons can also play important roles, for example in diffuse
interstellar clouds. Rate coefficients for collisions with H$_2$ based
on the potential energy surface (PES) of Jankowski \& Szalewicz (1998,
hereafter JS98) have been calculated recently
\citep{mengel01,flower01}. These rates were found to be in reasonable
agreement (typically within a factor of 2 at 10~K) with those computed
on older and less accurate PES (e.g. \citet{schinke85}). However, in
contrast to previous works, \citet{mengel01} and \citet{flower01}
found the inelastic rates with $\Delta J=1$ larger than those with
$\Delta J=2$ for para-H$_2$. This result reflects the crucial
importance of the PES, whose inaccuracies are one of the main sources
of error in collisional rate calculations, especially at low
temperatures (see, e.g., \citet{bala02}). The JS98 surface has been
checked against experiment by a couple of
authors. \citet{gottfried01}, in particular, computed second virial
coefficients below 300~K and showed that the attractive well of the
PES was slightly too deep. A modification which consists of
multiplying the PES by a factor of 0.93 was therefore suggested by
these authors. This modification was applied by \citet{mengel01} and
it was indeed found to produce better agreement with measurements of
pressure broadening cross sections. On the other hand, the original
PES was found to give a very good agreement with experimental
state-to-state rotational cross sections measured near 1000~cm$^{-1}$
\citep{antonova00}. At these energies, however, only the short-range
repulsive region of the PES is probed. Finally, \citet{flower01}
showed that the scaling of 0.93 has only minor influence on the
rotational rates and he employed the original PES.
  
Very recently, a new PES has been computed by \citet{jankowski05}
(hereafter JS05). In contrast to the JS98 surface, which was
calculated within the symmetry-adapted perturbation theory (SAPT), the
JS05 surface has been obtained using the supermolecular approximation
based on the CCSD(T) method using correlation-consistent basis sets
complemented by bonding functions and basis set extrapolation. It
turned that the major inaccuracy of JS98 was due to the use of a
correction for the effects of the third and higher orders
extracted in an approximate way from supermolecular SCF
calculations. It recently became clear that this correction should not
be used for nonpolar or nearly nonpolar monomers. If this term is
removed from JS98, the agreement of the resulting surface with JS05 is
excellent. \citet{jankowski05} have employed their new PES to
calculate rovibrational energy levels of the para-H$_2-$CO complex and
they obtained a very good agreement (better than 0.1~cm$^{-1}$
compared to 1~cm$^{-1}$ for the JS98 surface) with the measurements by
\citet{mckellar98}. The computed values of the second virial
coefficients also agree very well with the experimental ones. Full
details can be found in \citet{jankowski05}. In the present paper, we
present new rate constants for the rotational excitation of CO by
H$_2$ based on the JS05 surface. Our results are restricted to low
temperatures (T$<$70~K) because the new PES calculations are found to
affect the rotational cross sections only at low collision energy
(E$_{\rm coll}<$ 60~cm$^{-1}$). As shown below, differences with the
results of \citet{flower01} range from a few percent to 50~$\%$. Such
effects, if not dramatic, are significant in view of the special
importance of the CO$-$H$_2$ system in astrophysics. Furthermore, the
present results are of particular interest for comparisons with future
measurements planned in Rennes by Sims and collaborators
\citep{sims05}.

%
  
\section{Method}
\subsection{Potential Energy Surface}

Following the conventions of JS98, the CO$-$H$_2$ ``rigid-body'' PES
is described by the following four Jacobi coordinates: three angles
$\theta_1$, $\theta_2$ and $\phi$ and the distance $R$ denoting the
intermolecular centre of mass separation. The angles $\theta_1$ and
$\theta_2$ are the tilt angles of H$_2$ and CO with respect to the
intermolecular axis, and $\phi$ is the dihedral angle. In the linear
geometry, the oxygen atom points toward H$_2$ when $\theta_2$=0. The
JS05 surface has been obtained by averaging over the intramolecular
ground state vibration of H$_2$, while CO was treated as rigid at its
vibrational ground-state averaged geometry. The JS05 surface has a
global depth of -93.049~cm$^{-1}$ for the linear geometry with the C
atom pointing toward the H$_2$ molecule ($\theta_1=0$,
$\theta_2$=180$^{\rm o}$, $\phi=0$) and for a centre of mass
separation $R_{\rm min}=$7.92~bohrs.
\begin{figure}
  \resizebox{\hsize}{!}{\includegraphics{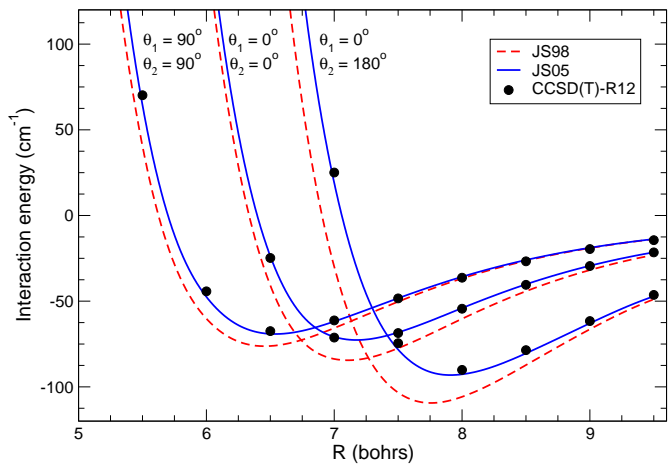}}
  \caption{H$_2-$CO interaction energy as a function of $R$ for three
    coplanar configurations: $(\theta_1, \theta_2)$= ($90^{\rm o}$,
    $90^{\rm o}$), ($0^{\rm o}$, $0^{\rm o}$) and ($0^{\rm o}$,
    $180^{\rm o}$). The dashed and solid lines denote the JS98 and
    JS05 calculations, respectively.  The filled circles correspond to
    independent CCSD(T)-R12 calculations (see text).}
\end{figure}
The (unscaled) JS98 and JS05 surfaces are
presented in Fig.~1 and compared to independent
CCSD(T)-R12 calculations using ground-state averaged geometries 
for CO and H$_2$. 

The latter calculations can be considered as high accuracy
state-of-the-art {\it ab initio} calculations.  Indeed the CCSD(T)-R12
approach\footnote{The R12 coupled cluster theory with singles, doubles
and perturbative triples properly describes the electron-electron cusp
by including explicitly the inter-electronic coordinates into the
wavefunction (\citet{noga94}, see also \citet{noga02} for a review.)}
has been shown to produce intermolecular energies with near
spectroscopic accuracy without recourse to extrapolation
\citep{halklo99,muller00,faure05}. In order to save some computing
time and to improve the numerical stability of the R12 operator, the
core orbitals were frozen\footnote{ This approximation is of little
consequence as core correlation effects are minor for the CO-H$_2$
interaction (about 0.3 cm$^{-1}$ at 7 bohrs separation).}  in the
CCSD(T)-R12 calculations.  We employed R12-suited basis sets including
up to $g$ functions for C and O, and $d$ functions for H (as developed
in \cite{cccc03, kedz05}). The R12 results are assumed to approach the
basis set limit within $1-2$ cm$^{-1}$. In particular the inclusion of
a large set of bonding functions (corresponding to the R12-suited H
set) affects the interaction by $\approx$ 0.5 cm$^{-1}$ only at 7
bohrs separation, while the counterpoise corrections remain very
small, generally below 1 cm$^{-1}$.

Here, we can observe that the JS05 surface agree very well with the
CCSD(T)-R12 results while the JS98 surface overestimates the depth of
the CO$-$H$_2$ potential well by about 10$-$20~$\%$, as anticipated by
\citet{gottfried01}. However, the JS05 surface lies slightly below the
CCSD(T)-R12 one. This agrees with the observation in JS05, based on
extrapolations involving bases up to quintuple zeta quality, that the
potential is about 1-2 cm$^{-1}$ too deep in the region of the
global minimum.  On the other hand, the comparison of the virial
coefficients from JS05 with experiment suggests that the exact
potential is still deeper, probably due to the effects resulting from
theory levels beyond CCSD(T). We also checked that the R12 values are
recovered within 1 cm$^{-1}$ by a basis set extrapolation involving
larger (triple and quadruple zeta) sets and the popular bonding
functions proposed by \citet{wmsj95}. These comparisons provides a
further check of the quality of the new CO$-$H$_2$ PES and also
confirm the suspected inaccuracies of the JS98 surface.

In order to interface the JS98 and JS05 surfaces with the scattering
code, the \textsc{fortran} routines of \citet{jankowski98,jankowski05}
were employed to generate a grid of 57,000 geometries. These grid
points were distributed in 19 fixed intermolecular distances $R$ (in
the range 4$-$15~bohrs) via random sampling for the angular
coordinates. At each intermolecular distance, the interaction
potential $V$ was least-squares fitted over the following angular
expansion:

\begin{equation}
V(R,\theta_1, \theta_2, \phi) = \sum_{l_1 l_2 l} v_{l_1 l_2
l}(R) s_{l_1 l_2 l}(\theta_1, \theta_2, \phi),
  \end{equation}
where the basis functions $s_{l_1 l_2 l}$ are products of spherical
harmonics and are explicited in Eq.~(A9) of
\citet{green75}. These functions form an orthonormal basis set, and
the expansion coefficients $v_{l_1 l_2 l}$ can thus be written as 
\footnote{$\star$ denotes the complex conjugate. Note that the angular
conventions of \citet{green75} are slightly different from those of
\citet{jankowski98}.}
\begin{equation}
  v_{l_1 l_2 l}(R)= \int_{\theta_1} \int_{\theta_2} \int_{\phi}
  V(R,\theta_1,\theta_2,\phi) s_{l_1 l_2 l}^\star
  (\theta_1,\theta_2,\phi) \sin\theta_1 d\theta_1 \sin\theta_2
  d\theta_2 d\phi
\end{equation}
The latter expression remains valid for any truncated expansion.

We selected an initial angular expansion including all anisotropies up
to $l_1$=10 for CO and $l_2$=6 for H$_2$, resulting in 142 $s_{l_1 l_2
l}$ functions.  The accuracy of the angular expansion was monitored
using a self-consistent Monte Carlo error estimator \citep{risttbs},
which also permitted to select the most pertinent angular terms. The
resulting set was composed of 60 basis functions involving
anisotropies up to $l_1=7$ and $l_2=4$. A cubic spline interpolation
was finally employed over the whole intermolecular distance range and
was smoothly connected with standard extrapolations to provide
continuous expansion coefficients suitable for scattering
calculations. The final accuracy of this 60 term expansion was found
to be better than 1~cm$^{-1}$ in the global minimum region of the PES
while the individual expansion coefficients were converged within 0.01
cm$^{-1}$. It should be noted that the expansion does not reproduce
exactly the values of the potential energy at the grid
points\footnote{The residual differences indicate the cumulative
effects of the higher expansion terms.}, in contrast to the approach
followed by \citet{flower01} who used the same number (25) of angular
geometries and basis functions. The main advantage of our approach is
to guarantee the convergence of the individual expansion coefficients,
in particular in the highly anisotropic short-range part ($R<6$~bohrs)
of the interaction.

\subsection{Scattering calculations}
  
The \textsc{molscat} code \citep{molscat} was employed to perform the
rigid-rotor scattering calculations reported below. All calculations
were made using the rigid rotor approximation, with rotational
constants $B_{\rm CO}=1.9225$ cm$^{-1}$ and $B_{\rm H_2}=60.853$
cm$^{-1}$. For para-H$_2$ ($j_2=0$), calculations were carried out at
total energies (collision plus CO rotation) ranging from 4 to
520~cm$^{-1}$. Full close-coupling calculations were performed between
4 and 160~cm$^{-1}$, with steps of 0.2~cm$^{-1}$ in the resonance
range 4$-$120~cm$^{-1}$, and a much coarser step of 5~cm$^{-1}$
between 120 and 160~cm$^{-1}$. Between 160 and 520~cm$^{-1}$, the
coupled-state approximation \citep{mcguire74} was shown to reproduce
full close-coupling calculations within a few percent and was thus
employed. The energy step was 20~cm$^{-1}$ between 160 and
340~cm$^{-1}$, and 30 cm$^{-1}$ at higher energies. For ortho-H$_2$,
the same collision energy grid was used, resulting in total energies
in the range 126$-$642~cm$^{-1}$. The propagator used was the
log-derivative propagator \citep{alexander87}. Parameters of the
integrator were tested and adjusted to ensure a typical precision of
1-2 \%. All the calculations with para-H$_2$ used a $j_{2}=0, 2$
basis, while the ortho-H$_2$ calculations used only a $j_{2}=1$
basis. Tests of this basis were made at the highest energy, where
inclusion of the $j_{2}=3$ (closed) state was found to affect cross
sections by less than one percent. At least three closed channels were
included at each energy for the CO rotational basis. At the highest
energy, we had thus a $j_{1}\le 18$ basis. Transitions among all
levels up to $j_1=5$ were computed. Finally, rate constants were
calculated for temperatures ranging from 5 to 70~K by integrating the
cross sections over essentially the whole collision energy range
spanned by the corresponding Maxwell-Boltzmann distributions. The
energy grid was chosen so that the highest collision for all
transitions be about ten times larger than the highest
temperature. The accuracy of the present rates is expected to lie
between 5 and 10~\% at 70K, and probably better at lower temperatures.
%

\section{Collisional cross sections and rates}

\begin{figure} 
    \resizebox{\hsize}{!}{\includegraphics{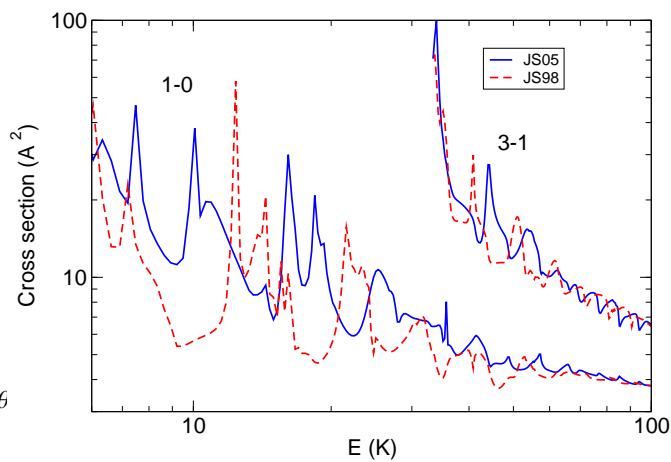}}
    \caption{Inelastic deexcitation cross sections for CO in
    collisions with para-H$_2$ as a function of total energy. Both
    curves are the results of our calculations with the unscaled JS98
    PES (dashed curves), and the JS05 PES (solid curves).}
    \label{cross}
  \end{figure}
  
Figure \ref{cross} clearly illustrates the fact that the resonance
pattern, which is here of 'shape' rather than 'Feshbach' type as
discussed by \citet{flower01}, strongly depends on the interaction
potential. We also note that the very fine grid used at the lowest
energies was necessary for a proper description of the resonances. It
can be observed that the cross sections converge at higher energies
(above the resonance regime), as the collision energy becomes large
compared to the potential energy well depth. As a result, rate
constants are expected to differ significantly from those of
\citet{flower01} at very low temperatures only.

\begin{table}[h!]
  \caption{Rotational deexcitation rates of CO in collision with
    para-H$_2$ ($j_2=0$) at three temparatures.  I and F are the CO
    initial and final rotation states, respectively. All rates are
    given in units of $10^{-10}$ cm$^3$/s. Third column recalls the
    rates of \citet{flower01}, obtained with the unscaled JS98
    surface. Fourth column presents our rates obtained with the JS05
    surface. In the 30~K panel, the rates obtained with our
    expansion of the JS98 surface are also presented.}
  \label{para}
  \centering
  \begin{tabular}{c c c c} 
    \hline\hline
    I & F & Flower & This work \\
    \hline
    &&Temp.: 10 K \\
    1 &   0 &      0.281 &      0.332 \\
    2 &   0 &      0.229 &      0.305 \\
    3 &   0 &      0.064 &      0.052 \\
    4 &   0 &      0.020 &      0.031 \\
    5 &   0 &      0.007 &      0.010 \\ 
    2 &   1 &      0.697 &      0.722 \\
    3 &   1 &      0.412 &      0.502 \\
    4 &   1 &      0.114 &      0.096 \\
    5 &   1 &      0.042 &      0.062 \\
    3 &   2 &      0.816 &      0.794 \\
    4 &   2 &      0.526 &      0.620 \\
    5 &   2 &      0.153 &      0.125 \\
    4 &   3 &      0.888 &      0.822 \\
    5 &   3 &      0.656 &      0.759 \\
    5 &   4 &      0.925 &      0.662 \\
    \hline  
  \end{tabular}
  \begin{tabular}{c c c c} 
    \hline\hline
    I & F & Flower & This work \\
    \hline
    &&Temp.: 70 K \\
    1 &   0 &      0.336 &      0.344 \\ 
    2 &   0 &      0.300 &      0.320 \\ 
    3 &   0 &      0.098 &      0.083 \\ 
    4 &   0 &      0.035 &      0.044 \\ 
    5 &   0 &      0.020 &      0.024 \\ 
    2 &   1 &      0.636 &      0.602 \\ 
    3 &   1 &      0.486 &      0.509 \\ 
    4 &   1 &      0.177 &      0.161 \\ 
    5 &   1 &      0.061 &      0.080 \\ 
    3 &   2 &      0.681 &      0.659 \\ 
    4 &   2 &      0.567 &      0.591 \\ 
    5 &   2 &      0.221 &      0.205 \\ 
    4 &   3 &      0.678 &      0.689 \\ 
    5 &   3 &      0.631 &      0.654 \\ 
    5 &   4 &      0.748 &      0.705 \\ 
    \hline    
  \end{tabular}
 \begin{tabular}{c c c c c} 
    \hline\hline
    I & F & Flower & This work JS98 & This work JS05 \\
    \hline
    &&Temp.: 30 K \\
    1 &   0 &   0.300 &    0.288 &    0.327 \\
    2 &   0 &   0.269 &    0.297 &    0.315 \\
    2 &   1 &   0.661 &    0.613 &    0.612 \\
    3 &   0 &   0.072 &    0.059 &    0.060 \\
    3 &   1 &   0.456 &    0.492 &    0.509 \\
    3 &   2 &   0.695 &    0.677 &    0.670 \\
    4 &   0 &   0.026 &    0.029 &    0.037 \\
    4 &   1 &   0.130 &    0.110 &    0.118 \\
    4 &   2 &   0.555 &    0.604 &    0.612 \\
    4 &   3 &   0.747 &    0.716 &    0.712 \\
    5 &   0 &   0.010 &    0.010 &    0.014 \\
    5 &   1 &   0.047 &    0.054 &    0.068 \\
    5 &   2 &   0.175 &    0.149 &    0.160 \\
    5 &   3 &   0.648 &    0.712 &    0.705 \\
    5 &   4 &   0.816 &    0.740 &    0.713 \\
    \hline    
  \end{tabular}
\end{table}
\begin{table}
  \caption{Rotational deexcitation rates of CO in collision with
    ortho-H$_2$ ($j_2=1$). The conventions are the same as for the
    para case.}
  \label{ortho}
  \centering
  \begin{tabular}{c c c c c} 
    \hline\hline
    I & F & Flower & This work \\
    \hline
    &&Temp.: 10 K \\
    1 &   0 &      0.388 &      0.379  \\ 
    2 &   0 &      0.596 &      0.559  \\ 
    3 &   0 &      0.105 &      0.071  \\ 
    4 &   0 &      0.083 &      0.061  \\ 
    5 &   0 &      0.015 &      0.016  \\ 
    2 &   1 &      0.730 &      0.712  \\ 
    3 &   1 &      0.842 &      0.831  \\ 
    4 &   1 &      0.159 &      0.124  \\ 
    5 &   1 &      0.124 &      0.105  \\ 
    3 &   2 &      0.801 &      0.748  \\ 
    4 &   2 &      0.862 &      0.894  \\ 
    5 &   2 &      0.188 &      0.150  \\ 
    4 &   3 &      0.827 &      0.786  \\ 
    5 &   3 &      0.896 &      0.972  \\ 
    5 &   4 &      0.871 &      0.695  \\ 
    \hline 
  \end{tabular}
  \begin{tabular}{c c c c c} 
    \hline\hline
    I & F & Flower & This work \\
    \hline
    &&Temp.: 70 K \\
    1 &   0 &      0.402 &      0.350 \\ 
    2 &   0 &      0.629 &      0.468 \\ 
    3 &   0 &      0.114 &      0.090 \\ 
    4 &   0 &      0.079 &      0.063 \\ 
    5 &   0 &      0.027 &      0.028 \\ 
    2 &   1 &      0.684 &      0.620 \\ 
    3 &   1 &      0.858 &      0.728 \\ 
    4 &   1 &      0.199 &      0.174 \\ 
    5 &   1 &      0.124 &      0.111 \\ 
    3 &   2 &      0.732 &      0.686 \\ 
    4 &   2 &      0.871 &      0.808 \\ 
    5 &   2 &      0.236 &      0.216 \\ 
    4 &   3 &      0.757 &      0.711 \\ 
    5 &   3 &      0.880 &      0.844 \\ 
    5 &   4 &      0.779 &      0.719 \\ 
    \hline
  \end{tabular}
\end{table}

We present in Tables \ref{para} and \ref{ortho} our rate constants at
10, 30 and 70~K, and those of \citet{flower01} for comparison. We note
that especially at 10~K, rates can vary up to 50\%. These differences
reflect the influence of three independent parameters: (i) the high
accuracy of the new {\it ab initio} potential energy calculations,
(ii) the use of a well converged expansion of the PES over the angular
basis functions and (iii) the careful description of resonances. It is
however not clear how to discriminate among the three effects. At 70K
(and beyond), our rates and those of \citet{flower01} agree within
typically 10$-$20~\%, as anticipated from the convergence of the cross
sections at high energy. We therefore conclude that the differences
between our results and those of \citet{flower01} are astrophysically
relevant for temperatures below 70~K. Above this value, the
inaccuracies of the JS98 PES have only a minor influence on the
rotational rates and the results of \citet{flower01} and
\citet{mengel01} are reliable.

The para-H$_2$ rates presented in the 30~K panel in Table \ref{para}
illustrate the respective contribution of the surface and of the
subsequent angular expansion and collisional treatment. Data in
columns 1 and 2 are based upon the same JS98 surface and the
differences are ascribable to the better convergence of the angular
expansion and to the finer description of the resonances.  Conversely,
data in columns 2 and 3 are based upon the same angular expansion and
collisional treatment, and the differences only reflect the
improvement in the surface. Depending on initial and final rotation
states, various situations may occur. In some cases there is a
compensation of errors for the surface (as for 2-1, 3-0, 3-2), or for
the expansion and collisional treatment (as for 5-0). Collisional and
surface errors may also partially cancel out, as for the 1-0 rate, or
add up as for the 2-0 rate. While the detailed behaviour seems complex
and irregular, in average the errors introduced by the older JS98
surface are of the same order of magnitude as the errors introduced by
the limited angular expansion and energy mesh in the older work by
\citet{flower01}.

It is interesting to note that for collisions with para-H$_2$,
inelastic rates with $\Delta j_1=1$ are larger than those with $\Delta
j_1=2$ while the reverse applies for ortho-H$_2$. The propensity to
favor even $\Delta j_1$ over odd $\Delta j_1$ has been explained
semiclassically by \citet{miller77} in terms of an interference effect
related to the even anisotropy of the PES. The reverse propensity can
also occur if the odd anisotropy is sufficiently large, as observed
experimentally in the case of CO-He \citep{carty04}. Here, the
propensity depends on the quantum state of the projectile because in
the para case ($j_2=0$), some terms of the PES expansion (those
associated with the quadrupole of H$_2$) vanish
identically. Therefore, the even anisotropy of the PES is larger with
ortho-H$_2$ than with para-H$_2$ and $\Delta j_1=2$ are favored. Note,
however, that these propensity rules depend on $j_1$, $\Delta j_1$ and
the temperature. Another interesting point is that the differences
with the results of \citet{flower01} are generally smaller with
ortho-H$_2$ than with para-H$_2$. This again reflects the different
anisotropies of the PES for para and ortho-H$_2$: in the ortho case,
cross sections are larger and scattering calculations are less
sensitive to small changes in the PES. Similar effects have also been
observed recently in the case of H$_2$O-H$_2$ \citep{dubernet05}.

It is also of some interest to compare the present results with the
CO-He rates obtained by \citet{dalgarno02}. It is generally assumed
that rates with para-H$_2$ ($j_2=0$) should be about 50\% larger than
He rates owing to the smaller collisional reduced mass and the larger
size of H$_2$. We have checked that the present H$_2$ ($j_2=0$) rates
are in fact within a factor of 1$-$3 of the He rates. Thus, as already
shown by \citet{phillips96} in the case of H$_2$O, rates for
excitation by H$_2$ are not adequately represented by scaled rates for
excitation by He.

Finally, for use in astrophysical modelling, the temperature
dependence of the above transition rates $k(T)$ have been fitted by
the analytic form used by \citet{bala99}:
\begin{equation}
  \log_{10}k(T)=\sum_{n=0}^4 a_nx^n, 
  \label{fitrates}
\end{equation}
where the $x$ factor was changed from $T^{-1/3}$ to $T^{-1/6}$ to
achieve a better fitting precision. The quadratic error on this fit is
lower than 0.1\% for all transitions. The coefficients of this fit are
provided as online material. For those who plan to use them, we
emphasize that these fits are only valid in the temperature range 5
$\leq T\leq 70$~K. Finally, excitation rates are not given in
this paper, but can be obtained by application of the detailed
balance.

%
  
\section{Conclusion}

Cross sections for the rotational (de)excitation of CO by ground state
para- and ortho-H$_2$ have been computed using quantum scattering
calculations for collision energies in the range 1$-$520~cm$^{-1}$. A
new, highly accurate, CO$-$H$_2$ potential energy surface has been
employed and it has been shown to strongly influence the resonance
structure of the cross sections in the very low collision energy
regime ($E_{\rm coll} \lesssim 60$~cm$^{-1}$). Conversely, at higher
energies, the effect of the new potential was found to be only
minor. As a result, the present rate constants are found to differ
significantly from those of \citet{flower01}, obtained on a previous
and less accurate potential, only for temperatures lower than
70~K. Transitions among all levels up to $j_1=5$ only were computed as
higher states are generally not populated at the low temperatures
investigated here.

This work illustrates the relationship between the accuracy of the
potential energy surface and the accuracy of the corresponding
inelastic cross sections for a simple system of astrophysical
relevance. Even at low temperatures, moderate inaccuracies in the
surface (in the 10-20\% range) result in semi-quantitative inelastic
rates with typical errors in the 20-50\% range with no dramatic
amplification of errors. However, this optimistic conclusion presents
also a harder counterpart, as we show that in order to obtain accurate
inelastic rates one has to satisfy all three conditions, i.e., use an
accurate surface, a properly converged angular expansion, and a
detailed description of the cross-section resonances.
Consequently, the improvement of the accuracy of inelastic rates,
especially at higher temperatures when the number of channels is
large, represents a considerable computational effort. In this
respect, the work by \citet{flower01} represented an excellent
compromise to achieve a typical 20\% accuracy below 100~K.

This work extends the initial objectives of the ``Molecular Universe''
European FP6 network (2005-2008) and participates to the current
international efforts to improve the description of the underlying
microscopic processes in preparation of the next generation of
molecular observatories.

%

\begin{acknowledgements}
  
  CCSD(T)-R12 calculations were performed on the IDRIS and CINES
  French national computing centers (projects no. 051141 and x2005 04
  20820).  MOLSCAT calculations were performed on an experimental
  cluster using a subset of the computer grid tools (under project
  Cigri of the ``Action Incitative GRID'') with the valuable help from
  F. Roch and N. Capit.  This research was supported by the CNRS
  national program ``Physique et Chimie du Milieu Interstellaire'' and
  the ``Centre National d'Etudes Spatiales''. MW was supported by the
  Minist\`ere de l'Enseignement Sup\'erieur et de la Recherche. KS
  acknowledges the NSF CHE-0239611 grant.

\end{acknowledgements}
  
\bibliography{corr_proof}

\Online
\appendix
\begin{table}
  \begin{tabular}{ccccccc}
    $I$ & $F$ & $a_0$ & $a_1$ & $a_2$ & $a_3$ & $a_4$ \\
    \hline
    1 & 0 & -10.413 & 4.162 & -20.599 & 32.563 & -16.889 \\
    2 & 0 & -9.202 & -7.997 & 18.452 & -18.717 & 6.910 \\
    2 & 1 & -0.379 & -59.429 & 131.662 & -127.161 & 45.605 \\
    \hline
    3 & 0 & -3.319 & -35.329 & 53.745 & -29.641 & 2.471 \\
    3 & 1 & -7.042 & -20.755 & 48.788 & -49.875 & 18.607 \\
    3 & 2 & -1.976 & -48.986 & 107.139 & -102.134 & 36.288 \\
    \hline
    4 & 0 & -8.004 & -16.164 & 28.359 & -21.986 & 6.026 \\
    4 & 1 & 1.979 & -69.801 & 140.821 & -125.047 & 40.556 \\
    4 & 2 & -5.102 & -34.014 & 82.717 & -87.211 & 33.665 \\
    \hline
    4 & 3 & -1.891 & -49.861 & 109.419 & -103.434 & 35.726 \\
    5 & 0 & 7.564 & -106.151 & 220.187 & -205.770 & 72.462 \\
    5 & 1 & -5.209 & -32.541 & 67.718 & -63.996 & 23.105 \\
    \hline
    5 & 2 & 4.736 & -91.653 & 203.779 & -201.295 & 73.681 \\
    5 & 3 & -5.607 & -31.380 & 78.077 & -83.702 & 32.930 \\
    5 & 4 & -0.510 & -60.401 & 137.793 & -134.344 & 46.449 \\

  \end{tabular}
  \caption{Fitting coefficients of CO-para-H$_2$ rates, following
  formula \ref{fitrates}. $I$ and $F$ are the initial and final
  rotation states, respectively. The rates thus obtained are in
  $cm^3/s$.}
\end{table}
\begin{table}
  \begin{tabular}{ccccccc}
    $I$ & $F$ & $a_0$ & $a_1$ & $a_2$ & $a_3$ & $a_4$ \\
    \hline
    1 & 0 & -7.351 & -11.062 & 5.033 & 16.394 & -14.176 \\
    2 & 0 & 0.331 & -68.556 & 162.123 & -167.243 & 63.863 \\
    2 & 1 & 4.370 & -90.464 & 207.213 & -207.995 & 77.557 \\
    \hline
    3 & 0 & 4.128 & -88.472 & 191.123 & -181.945 & 64.170 \\
    3 & 1 & -1.466 & -57.252 & 138.316 & -144.752 & 55.539 \\
    3 & 2 & -0.194 & -61.716 & 141.443 & -142.699 & 53.788 \\
    \hline
    4 & 0 & -6.118 & -31.232 & 70.967 & -70.685 & 25.956 \\
    4 & 1 & 5.330 & -96.359 & 215.047 & -212.299 & 77.662 \\
    4 & 2 & -2.940 & -49.481 & 124.615 & -135.311 & 53.609 \\
    \hline
    4 & 3 & -1.454 & -53.713 & 122.102 & -121.151 & 44.472 \\
    5 & 0 & 5.241 & -96.162 & 206.655 & -199.255 & 72.234 \\
    5 & 1 & -6.365 & -27.561 & 61.697 & -61.299 & 22.802 \\
    \hline
    5 & 2 & 6.689 & -107.753 & 249.875 & -256.409 & 97.478 \\
    5 & 3 & -3.812 & -45.669 & 119.845 & -134.502 & 54.860 \\
    5 & 4 & 1.082 & -71.954 & 169.486 & -173.174 & 64.402 \\
  \end{tabular}
  \caption{Fitting coefficients of CO-ortho-H$_2$ rates, following
  formula \ref{fitrates}. $I$ and $F$ are the initial and final
  rotation states, respectively. The rates thus obtained are in
  $cm^3/s$.}
\end{table}

\end{document}